\documentclass[aps,twocolumn,showpacs]{revtex4}
\usepackage{graphicx}
\usepackage{amssymb}
\usepackage{amsmath}
\newcommand{\be}{\begin{equation}}
\newcommand{\ee}{\end{equation}}
\newcommand{\ben}{\begin{eqnarray}}
\newcommand{\een}{\end{eqnarray}}
\newcommand{\bes}{\begin{subequations}}
\newcommand{\ees}{\end{subequations}}
\newcommand{\bb}{\bibitem}

\begin{document}
\title{First-order formalism for dark energy and dust}
\author{D. Bazeia,$^a$ L. Losano,$^a$ J.J. Rodrigues,$^a$ and R. Rosenfeld$\,^b$}
\affiliation{$^a$Departamento de F\'\i sica, Universidade Federal da Para\'\i ba, 58051-970 Jo\~ao Pessoa, Para\'\i ba, Brazil\\
$^b$Instituto de F\'\i sica Te\'orica, UNESP, Rua Pamplona, 145, 01405-900 S\~ao Paulo, Brazil}
\date{\today}

\begin{abstract}
This work deals with first-order formalism for dark energy and dust in standard cosmology, for models described by real scalar field in the presence of dust
in spatially flat space. The field dynamics may be standard or tachyonic, and we show how the equations of motion can be solved by first-order differential equations. We investigate a model to illustrate how the dustlike matter may affect the cosmic evolution using this framework.
\end{abstract}

\pacs{98.80.Cq}

\maketitle

{\bf I. Introduction.} The discovery of cosmic acceleration \cite{acce} has brought to the stage one of the greatest conundrums in modern cosmology. This so-called dark energy, the mysterious source of this acceleration, has to make up a very significant portion of the total energy density of the universe today \cite{reviews}. Recent observations \cite{obser} indicate that a simple cosmological constant of the appropriate size is still in good agreement with all data. However, this solution requires a large amount of fine tuning in order for the cosmological constant to dominate the energy density of the universe quite recently, the so-called coincidence problem. One widely studied possibility is that the cosmic acceleration may be fueled by quintessence \cite{qui}, a homogeneous fluid with time-dependent equation of state, which in standard Friedmann-Robertson-Walker (FRW) cosmology is described by real scalar fields.

In a very recent work, a method for investigating models of real scalar fields in spherical, flat or hyperbolic FRW background was proposed which uses Einstein's equation and the equation of motion for the scalar field in a very direct way \cite{bglm}. In this method, one follows a very specific route, using the potential of the scalar fields to infer how the scale factor or, alternatively, Hubble's parameter evolves in time. The power of the method developed is related to an important simplification, which leads to models governed by scalar field potential of very specific form, in general depending on two new functions, $W=W(\phi)$ and $Z=Z(\phi),$ which allows reducing the equations of motion to first-order differential equations. The presence of first-order equations was also explored in Ref.~{\cite{l}} in spatially flat FRW cosmology, and in other contexts in the recent Refs.~{\cite{abl,str}}. The methodology introduced in \cite{bglm} for FRW cosmology was shown to work for bent brane in Ref.~{\cite{abl}}. Together with Ref.~{\cite{str}}, some new issues concerning the brane/cosmology correspondence
have been brought and studied in several recent investigations \cite{BC}.

In the present work, our main motivation is to extend the first-order formalism of Ref.~{\cite{bglm}} to the important case which includes dustlike matter, that is, pressureless non-relativistic matter which can be of baryonic or dark nature.  The inclusion of dust is of current interest, and would set an important step towards more realistic cosmological scenarios. The fact that dustlike matter is intrinsically non-relativistic has added no important objection to our framework, and we believe that this issue will perhaps contribute to the recent observation that the first-order formalism of \cite{bglm} seems to be not intrinsically connected with supersymmetry \cite{str}.

The explicit extension of the first-order formalism to include dustlike matter in flat space-time is given below in Sec. {\bf II}, where we consider scalar field driven by standard or tachyonic dynamics \cite{tachyon}. In Sec.~{\bf III} we investigate a model to illustrate how the dustlike matter may affect the cosmic evolution. Finally, in Sec.~{\bf IV} we present our ending comments.

{\bf II. Framework.} We follow the program set forward in \cite{bglm} and we study the case of a flat FRW geometry, that is, we consider the line element
\be
ds^2=dt^2-a^2(t)(dr^2+r^2d\Omega^2)
\ee
where $\Omega$ identifies the angular spatial portion of the flat space-time, and $a(t)$ is the scale factor. We will use $H={\dot a}/a$ to denote the Hubble parameter. The action for a real scalar field in this flat FRW background is given by
\be\label{model}
S=\int\,d^4x\;{\sqrt{-g}\;\left(-\frac14\,R+{\cal L(\phi,\partial_\mu\phi)}\right)}
\ee
where $\phi$ describes the scalar field and we are using ${4\pi G}=1.$ In general, the energy-momentum tensor is given by $T^\mu_{\;\;\nu}=(\rho,-p,-p,-p),$ where $\rho$ and $p$ represent the total energy density and pressure. We use Einstein's equation to get $H^2=2\rho/3$ and ${\ddot a}/{a}=-(\rho+3p)/3.$

The issue now is to include dustlike matter. We write the corresponding energy density in the form $\rho=\rho_\phi+\rho_d,$ with $\rho_d(a)={\bar\rho}/a^3,$ where $\bar\rho$ is real and positive constant, describing the energy density today ($a=1$) of the non-relativistic matter, assumed to be spread homogeneously and isotropically into the Universe. We also have $p=p_\phi$ since $p_d=0,$ e.g., the dustlike matter is pressureless. In this case, the above equations become
\bes\label{fe1}
\ben
H^2&=&\frac23\,\rho_\phi+\frac23 \frac{\bar\rho}{a^3}
\\
\frac{\ddot a}{a}&=&-\frac13\,(\rho_\phi+3p_\phi)-\frac13\frac{\bar\rho}{a^3}
\een
\ees

To further explore the subject, it is necessary to determine the dynamics of the scalar field. We do this below, where the scalar field is used to describe dark energy under the standard or tachyonic dynamics.

The equations of motion for the scalar field depend on ${\cal L}(\phi,\partial_\mu\phi),$ which may be of the standard form 
\be\label{sm}
{\cal L}=\frac12\partial_\mu\phi\partial^\mu\phi-V(\phi)
\ee
The energy density and pressure that account for the scalar field model are given by 
\be
\rho_\phi=\frac12\dot\phi^2+V,\;\;\;\;\;p_\phi=\frac12\dot\phi^2-V
\ee
Also, the equation of motion has the form
\be\label{em1}
\ddot\phi+3H\dot\phi+\frac{dV}{d\phi}=0
\ee

We use the energy density and pressure to rewrite Eqs.~(\ref{fe1}) as
\bes\label{em0}
\ben
H^2=\frac13{\dot\phi}^2+\frac23V+\frac23\frac{\bar\rho}{a^3}\label{em3}
\\
{\dot H}=-\dot\phi^2-\frac{\bar\rho}{a^3}\label{em2}
\een
\ees
Thus, the set of Eqs.~(\ref{em1}) and (\ref{em0}) constitutes the equations that we have to deal with in the case of scalar fields with standard dynamics in flat geometry in the presence of dust.

In the standard view in cosmology, since $a=a(t),$ $\phi=\phi(t),$ and $H=H(t),$ from Einstein's equation we need to have the potential as a function of time. However, from the equation of motion for the scalar field we have $V=V(\phi)$, which shows the potential as a function of the scalar field. Thus, to make these two views equivalent we need to view Hubble's parameter as a function of the scalar field too. Hence we write
\be\label{be0}
H=W+\alpha\bar{\rho}Z
\ee
where $W=W(\phi)$ and $Z=Z(\phi)$ are in principle arbitrary function of $\phi,$ and $\alpha$ is constant. This is a first-order differential equation for the scale factor. It allows obtaining another equation, which responds for the scalar field evolution in the form
\be\label{bek}
\dot{\phi}=-W_{\phi}-\beta\bar{\rho}Z_{\phi}
\ee
where $\beta$ is another constant, and we consider $\alpha\neq\beta.$ The equations of motion (\ref{em1}) and (\ref{em0}) and the above Eqs.~(\ref{be0}) and (\ref{bek}) imply that the potential has to have the form
\be
V=\frac32(W+\alpha\bar{\rho}Z)^2-\frac12(W_{\phi}+\beta\bar{\rho}Z_{\phi})[W_{\phi}+(2\alpha-\beta)\bar{\rho}Z_{\phi}]
\ee
The presence of the function $Z(\phi)$ adds a new constraint, which emerges consistently:
\be\label{cons1}
W_{\phi}Z_{\phi\phi}+W_{\phi\phi}Z_{\phi}+2\beta\bar{\rho}Z_{\phi}Z_{\phi\phi}-3\alpha\bar{\rho}ZZ_{\phi}-3WZ_{\phi}=0
\ee 

In the absence of dust, the above potential becomes $V=(3/2)W^2-(1/2)W_{\phi}^2,$ as expected \cite{bglm,str}. We then notice that in the first-order framework the presence of dust requires the appearance of new interactions, described by the functions $W(\phi)$ and $Z(\phi),$ and their derivatives, and mediated by the energy density necessary to describe the dustlike component. This information is of interest, since it shows that the presence of first-order equations depends on the way the dark energy and dust interact with one another.

We define ${\bar q}={\ddot a}a/{\dot a}^2$ as the acceleration parameter. In general it can be written as ${\bar q}=1+{\dot H}/H^2.$ Here we have
\be
\bar{q}=1-\frac{(W_{\phi}+\alpha\bar{\rho}Z_{\phi})(W_{\phi}+\beta\bar{\rho}Z_{\phi})}{(W+\alpha\bar{\rho}Z)^2}
\ee

We now turn attention to the case of a scalar field of tachyonic nature. The importance of tachyon in cosmology is inspired in String Theory, and it is interesting to see if one can extend the above formalism to the case where dark energy evolves under tachyonic dynamics \cite{tachyon}.
In this case, we have to change the Lagrange density to the new form
\be\label{td}
{\cal L}=-V(\phi)\sqrt{1-\partial_{\mu}\phi\partial^{\mu}\phi}
\ee
It follows that the energy density and pressure for the scalar field are now given by
\be\label{rho-p}
\rho=V\bigl/{\sqrt{1-{\dot{\phi}}^{2}}},\;\;\;\;\;p=-V\sqrt{1-{\dot{\phi}}^{2}}
\ee
Also, the time evolution for the scalar field is driven by
\be
\ddot{\phi}+(1-{\dot{\phi}}^{2})\left(3H\dot{\phi}+\frac{V_{\phi}}{V}\right)=0
\ee
In the presence of dustlike matter we have to replace Eqs.~(\ref{em0}) to the new ones
\bes
\ben
H^2=\frac23 \frac{V}{\sqrt{1-{\dot{\phi}}^{2}}}+\frac23 \frac{\bar{\rho}}{a^3}
\\
\dot{H}=-\frac{{\dot{\phi}}^{2}}{\sqrt{1-{\dot{\phi}}^{2}}}\,V - \frac{\bar{\rho}}{a^3}
\een
\ees

The presence of the square root makes the algebraic calculations much more intricate, and if we insist to obtain first-order equations in the tachyonic case, we realize that the choice $H=W+\alpha\bar\rho Z$ does not work anymore. However, a solution is possible if we write  
\be\label{tacnew}
H=\sqrt{W^2+\frac23 \alpha \bar{\rho}Z^2}
\ee
As before, $\alpha$ is constant and $Z=Z(\phi).$ In this case we have 
\be\label{phidot}
\dot{\phi}=-\frac23\,\frac{W_{\phi}}{W\sqrt{W^{2}+(2/3)\,\alpha\bar{\rho}Z^2}}
\ee
and the potential is given by
\be\label{tp}
V=\frac32\,W^2\left(1-\frac49\,\frac{W^2_\phi/W^2}{W^2+(2/3)\,\alpha\bar{\rho}Z^2}\right)^{1/2}
\ee
The two functions $W=W(\phi)$ and $Z=Z(\phi)$ now obey the consistency condition:
\be\label{cons}
W^{3}Z+\frac23 \alpha\bar{\rho}\,WZ^{3}-\frac49\,W_{\phi}Z_{\phi}=0
\ee

Energy density and pressure are now given by (\ref{rho-p}), and so the equation of state for the tachyonic model has the form $\omega=\dot\phi^2-1.$ Recent observations \cite{wtoday} impose the constraint $\omega\in[-1.10, -0.98],$ showing that $\dot\phi$ must be close to zero. If we consider $\dot\phi$ small we may get to the slow-roll approximation \cite{SR}. As it was already shown in Ref.~{\cite{padma}}, the slow-roll approximation changes the tachyonic dynamics back to the standard dynamics in flat spacetime. In the presence of dust, which is the case under consideration, the same arguments apply: for the tachyonic density (\ref{td}) we can write \cite{MZ} $-V(\phi)(1-\dot\phi^2)^{1/2}\approx V(\phi)\dot\phi^2/2-V(\phi)=\dot\chi^2/2-V(\chi),$ where we have used the field redefinition $\chi=f(\phi).$ This reasoning leads to the nice result that the experiments make the above first order framework for tachyonic dynamics unnecessary.

{\bf III. Illustration.} In view of the above results, let us study a model of scalar field with standard dynamics in detail. As mentioned before, we work in the context of a flat universe, with total density given by the critical density $\rho_c$. Although it is possible to keep $\alpha$ and $\beta$ arbitrary, with $Z$ and $W$ consistently obeying the constraint \eqref{cons1}, in the present work we choose $\alpha=0$ and $Z=W,$ for simplicity. In this case the constraint leads to $2(1+\beta\bar{\rho})W_{\phi\phi}-3W=0,$ yielding $W=A\cosh(B\phi),$ with $A$ constant and $B=\pm\sqrt{3/2(1+\beta\bar{\rho})}.$ The resulting potential is given by
\be
V=\frac34 A^2\left((1+\beta\bar{\rho})\cosh^2(B\phi)+(1-\beta\bar{\rho})\right)
\ee
In this case we get for the field evolution
\be
\phi(t)=\frac1B\ln\left(\tanh\left(\frac34\,At\right)\right)
\ee
and for the scale factor
\be
a(t)=\left(-\frac2{3\beta A^2}\sinh^2\left(\frac32\,At\right)\right)^{1/3}
\ee

The Hubble parameter can be written as
\be
H(t)=A\tanh^{-1}\left(\frac32\,At\right)
\ee
and the total energy density has the form
\be
\rho(a)=-\;\frac{\;\rho_d\;}{\beta\bar\rho}\left(1-\frac{f(a)}{1+\left(1-f(a)\right)^{1/2}}\right)^2
\ee
where $f(a)=\frac32\beta A^2 a^3$, and $\beta$ and $A$ are fixed by requiring that $\rho(a=1)=\rho_c=1$ and $\rho_\phi(a=0)=0.$ 

We plot below in Fig.~[1] $\Omega_{\phi}(a)=1-\Omega_d(a)=\rho_{\phi}/(\rho_{\phi}+\rho_d)$ and $\Omega_d(a)$ for $a$ in the interval $0\leq a\leq 2,$ with $a=1$ today.  The plots are depicted for the values $A=0.70,$ $\beta=-3.83,$ and $\bar\rho=0.26$. We have that $\Omega_{\phi}=0.74$ and $\Omega_d=0.26$ today by construction and they crossover in the past at $a=0.70.$

\begin{figure}[ht]
\includegraphics[{height=4cm,width=7.6cm}]{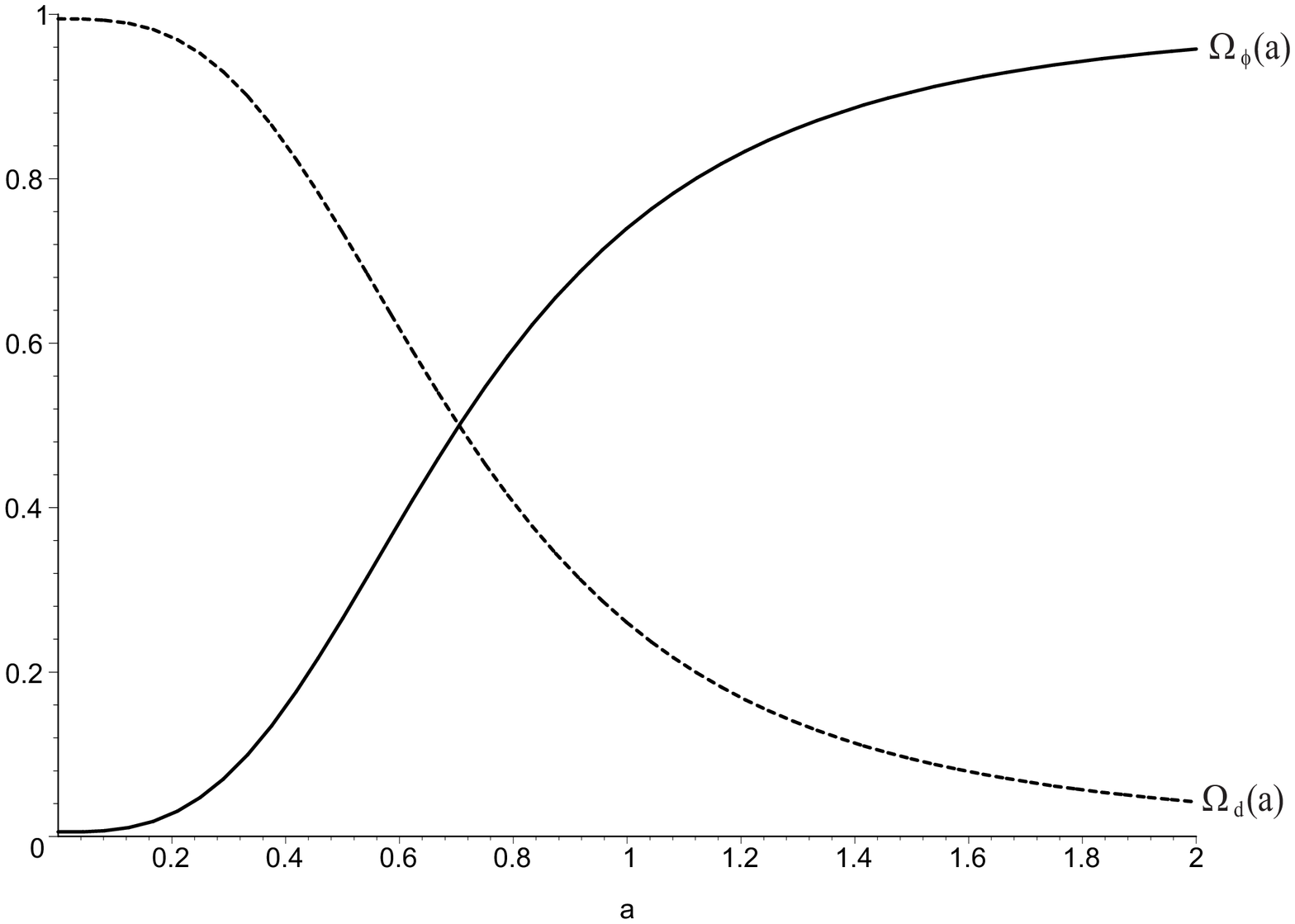}\hspace{.5cm}
\includegraphics[{height=4cm,width=7.6cm}]{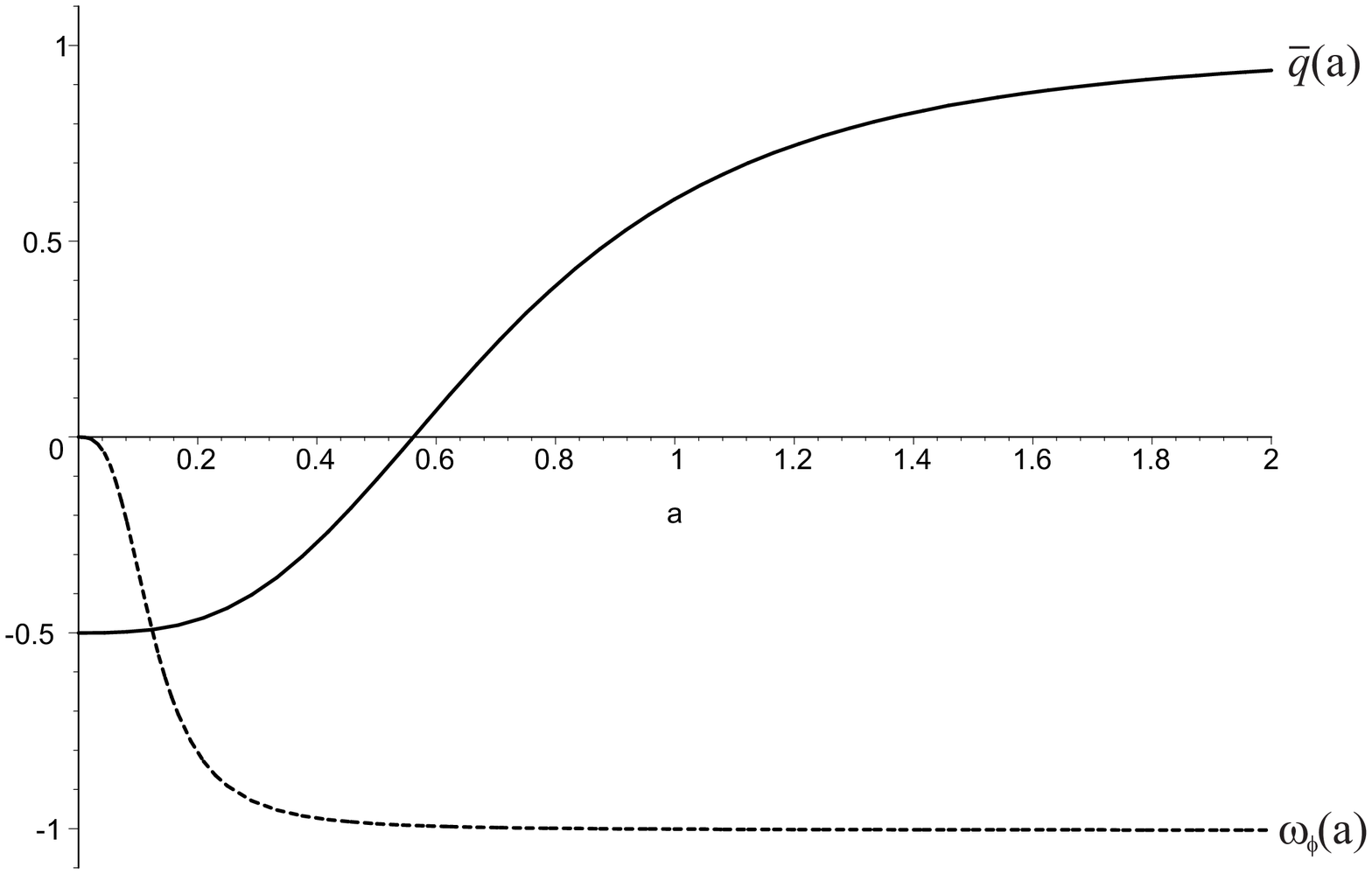}
\caption{Upper panel: plots of $\Omega_{\phi}$ (solid line) and $\Omega_d$ (dashed line). Lower panel: plots of the acceleration ${\bar q}$ (solid line) and the equation of state for the scalar field $\omega_\phi$ (dashed line). All the plots are done as a function of the scale factor $a$ in the interval $0\leq a\leq2,$ with $a=1$ today.}
\label{fig12}
\end{figure}

The acceleration parameter in this model is given by
\be
{\bar q}(t)=-\frac12+\frac32\,\tanh^2\left(\frac32\,At\right)
\ee
and can also be written as a function of $a,$ in the form 
\be
{\bar q}(a)=1-\frac32\frac1{1-f(a)}
\ee
To get the equation of state for the scalar field we recall that  $\omega_\phi=p_\phi/\rho_\phi,$ and so in the present model it can be written as
\be
\omega_\phi(a)=\frac{4f(a)}{(1+\beta\bar{\rho})\left(1+\left(1-f(a)\right)^{1/2}\right)^2+2\beta\bar{\rho}f(a)}
\ee

In Fig.~[1] we also show how the acceleration behaves as a function of the scale factor, for the same values of parameters used before. We notice that $\bar q$ goes to unit as $a$ increases to very large values. Also, it changes sign at $a=0.56,$ where $\Omega_{\phi}=0.34$ and $\Omega_d=0.66.$ In the same figure we also show how the equation of state $\omega_\phi(a)$ varies with the scale factor, $a.$ We have that at $a=0.70,$ where $\Omega_\phi=\Omega_d,$ the acceleration and equation of state are $\bar q=0.24$ and $\omega_\phi\approx-1.$

The results of the above model show good agreement with the observational data. In particular, the crossover between dark energy and dust occurs at $a=0.70,$ slightly above the expected value. The model illustrates the application of the first-order formalism, and it is particularly limited to recent epochs, where radiation is not important. It shows that the inclusion of dust makes the dynamics very interesting, being governed by first-order equations, with the time evolution changing from deceleration to acceleration at $a=0.56,$ for dark energy contributing with 34\% and dust with 66\% of the total energy density.

We can investigate other possibilities, with $\alpha\neq0$ and $\beta=0,$ and with $\alpha\neq0$ and $\beta\neq0,$ which lead to different relations between
$Z$ and $W,$ guided by the constraint \eqref{cons1}. In general, other models with polynomial and non-polynomial potentials can be introduced and will be further explored in another work, where we will also consider the presence of radiation. 

{\bf IV. Final comments.} In this work we have shown how to extend the first-order formalism of Ref.~{\cite{bglm}} to include dustlike matter in the presence of dark energy with standard or tachyonic dynamics. The crucial ingredient was the introduction of the new functions, $W=W(\phi)$ and $Z=Z(\phi).$ 

The standard and tachyonic models which we have investigated in the present work show that the formalism works nicely. They also illustrate 
the importance of finding first-order equations, which is directly related to the improvement of the process of searching for explicit solution.
In this sense, the present investigation is of direct interest to modern cosmology, and can be extended to several fields very naturally. A result of current interest is that dark energy and dust are naturally incorporated within the framework of Ref.~{\cite{bglm}}, and this compels us to ask whether it can be included in the supersymmetry-based formalism of Ref.~{\cite{str}}.

The authors would like to thank CAPES, CNPq, and PRONEX-CNPq-FAPESQ for partial support.


\end{document}